# Exploring wavefunction hybridization of magnon-magnon hybrid state

Bo Hu[1], Zong-Kai Xie[2], Jie Lu[3], and Wei He[2,] *

[1]Department of Physics, College of Sciences, Shanghai Maritime University, Shanghai 201306, China

[2]Beijing National Laboratory for Condensed Matter Physics, Institute of Physics, Chinese Academy of Sciences, Beijing 100190, China

[3]College of Physics Science and Technology, Yangzhou University, Yangzhou 225002, China

**Abstract:**

We investigate magnon–magnon hybrid states using a non-Hermitian two-band Hamiltonian and the concept of wavefunction hybridization. By comparing our model with micromagnetic simulations conducted on a synthetic antiferromagnet with strong magnon–magnon coupling, we successfully reproduce not only the resonance frequencies and linewidths but also the phases and amplitudes of the magnon wavefunction. The hybridization effect influences the dissipation rate, leading to the crossing of linewidths. Additionally, we quantify the magnon hybridization within a magnonic Bloch sphere, which enhances the ability to manipulate hybrid magnons for coherent information processing.

* Corresponding authors: hewei@iphy.ac.cn

# I. Introduction

The hybrid quantum system based on magnon has elicited significant attention due to its advancements in quantum information technologies [1–4]. Understanding this hybrid quantum system is crucial for the development of large-scale artificial quantum systems [5]. The use of magnon-based functional devices in quantum technologies has been encouraged. The hybridization of magnons with other quasiparticles, such as photons, phonons, and magnons themselves, introduces new hybrid quantum systems that can greatly enhance their utility in quantum information processing [2,6–9]. In recent years, magnon–magnon hybrid systems have attracted broad interest primarily due to the ease of achieving strong magnon–magnon coupling [10–13]. One of the most promising platforms for hybrid quantum systems is the synthetic antiferromagnetic system (SAF), which serves as a host for magnon–magnon coupling [9,14,15]. The dynamical excitations in SAF can be categorized into acoustic mode (AM) or optical mode (OM). Magnon–magnon coupling enables the coupling of these two magnon excitations, resulting in hybrid states. In the hybrid mode, the strength of magnon–magnon coupling determines the rate of coherent information exchange. Consequently, a strong magnon–magnon coupling is highly desirable. To date, most research on magnon–magnon coupling has focused on achieving a strong coupling strength, leading to the emergence of a characteristic anticrossing phenomenon between the two magnon modes. [16,17]

Understanding the magnon hybrid system cannot be solely achieved by measuring the level repulsion gap. The phenomenon of linewidth suppression or enhancement has been observed in magnon hybrid modes, and different explanations have been proposed depending on the specific hybrid system. In magnon–photon hybrid modes, it has been attributed to the two-mode hybridization [18–20]]. Conversely, in magnon–magnon hybrid modes, it has been interpreted as mutual spin pumping [21,22]. The effects of hybridization on linewidth in magnon–magnon hybrid modes still remain unclear. Recently, the dynamical phase shift between two magnets has been suggested as a parameter to differentiate the magnon hybrid mode in SAF. The dynamic phase shift at the hybrid mode is neither 0° nor 180° [23–25]. However, a universal theoretical framework is still lacking to explain and quantify this phase shift. To comprehensively explore magnon–magnon hybrid states, similar to atomic states described by wavefunctions, we propose the use of magnon wavefunctions to describe the magnon states. Notably, understanding and characterizing the wavefunction hybridization of magnon–magnon hybrid systems is crucial for coherent information encoding and processing. However, explicit studies focusing on magnon wavefunction hybridization in magnon–magnon hybrid systems are limited [26].

In this letter, we present a novel approach to explore the magnon–magnon hybrid state through the hybridization of magnon wavefunctions. To achieve this, we employ a non-Hermitian two-band Hamiltonian that allows us to determine the relevant parameters. The eigenvalues and eigenvectors obtained from our approach are in excellent agreement with the results obtained from a micromagnetic simulation conducted on a SAF. These results include the resonance frequency, linewidth, precession amplitude, and phase. The non-Hermitian component of our Hamiltonian is constructed based on the frequency–linewidth relationship, which captures the characteristic Gilbert damping behavior. We selected the wavefunctions associated with the AM and OM as our basis sets. The coupling between these modes leads to the formation of a new eigen wavefunction, which is a linear combination of the two basis states. The wavefunction hybridization induces a mixing of the linewidths, resulting in the suppression of one linewidth and the enhancement of the other. The observed phase behavior aligns with theoretical predictions, highlighting the non-Hermitian nature of the wavefunction hybridization. Furthermore, we investigate the evolution of magnon hybridization in a magnonic Bloch sphere as a function of an external field. By analyzing this evolution, we gain insights into the quantitative understanding and characterization of magnon hybridization for magnon–magnon hybrid systems.

## II. Theoretical Model

The phase shift alone is insufficient to fully characterize the hybridization. In addition to the phase shift, the weights of the AM and OM in a hybrid mode are also necessary. To address this, we construct a wavefunction $\emptyset = e^{i2\pi ft}(e^{i\delta}A_1\boldsymbol{m_1} + A_2\boldsymbol{m_2})$ to describe the dynamical magnetizations of the SAF. Here, $\boldsymbol{m_1}$ and $\boldsymbol{m_2}$ represent the time-dependent unit vectors of the top and bottom layer magnetizations, respectively. The wavefunction can be written more conveniently as $\emptyset = (e^{i\delta}A_1, A_2)^T$, where $T$ denotes the matrix transpose. The precession amplitudes are parameterized by $A_1$ and $A_2$ for the top and bottom layers, respectively. The phase $\delta$ is introduced to account for the relative phase shift between the two precessional magnetizations. In this context, $|a\rangle = (1, 1)^T$ ($\delta=0°$) and $|o\rangle = (-1, 1)^T$ ($\delta=180°$) present the eigenvectors of the pure AM and OM states, respectively. We illustrate the wavefunctions of pure AM and OM in Fig. 1(a) and consider them as the two basis wavefunctions $|a\rangle$ and $|o\rangle$, respectively. The wavefunction of the magnon–magnon hybrid mode can be obtained as a linear combination of $|a\rangle$ and $|o\rangle$, as depicted in Fig. 1(a).

We employ a non-Hermitian two-band Hamiltonian to fully characterize the hybrid mode by calculating the magnon energy and wavefunction. Specifically, we determine the eigenfrequencies and eigenvectors through an effective two-band Hamiltonian that incorporates coherent coupling between the magnons [27–29] as:

$$H = \begin{pmatrix} f_o - i\Delta f_o & g/2 \\ g/2 & f_a - i\Delta f_a \end{pmatrix}, \quad (1)$$

where $f_o$, $\Delta f_o$, $f_a$, and $\Delta f_a$, are the bare ferromagnetic resonance frequencies and linewidths for OM and AM, respectively. $g$ is the strength of the magnon–magnon coupling. The two-band Hamiltonian for magnon–magnon coupling was the work of Chai *et al.*, [28], but the effect of FMR linewidth was neglected in their work. Previous work based this non-Hermitian two-band Hamiltonian mostly focuses on the coupling-strength-dependent energy spectra. [29] In the current work, the diagonal of Eq. (1) is parameterized by the complex number to make the Hamiltonian non-Hermitian or the dynamical part of wavefunction as the manner of damped wave. We set $g$ as a real positive number, which means the absence of spin pumping. [22] After diagonalizing Eq. (1), the complex eigenfrequencies for two hybridized eigenmodes are obtained as [27]:

$$f_\pm = \tfrac{1}{2}(f_a + f_o) + \tfrac{i}{2}(\Delta f_a + \Delta f_o) \pm \tfrac{1}{2}\sqrt{\Delta^2 + g^2}, \quad (2)$$

where the complex parameter $\Delta = (f_a - f_o) - i(\Delta f_a - \Delta f_o)$. With the assumption $\alpha \ll 1$, the detected resonance frequency is $Re(f_\pm) = \tfrac{1}{2}(f_a + f_o) \pm \tfrac{1}{2}\sqrt{(f_a - f_o)^2 + g^2}$. Two branches of hybrid modes are classified as $f_+$ (upper) and $f_-$ (lower) branches. Also, we obtain the complex eigenvectors for the hybrid modes in the normalized solutions as (when $g \neq 0$):

$$\begin{pmatrix} A_o \\ A_a \end{pmatrix}_\pm = \frac{1}{\sqrt{g^2 + \left(\Delta \pm \sqrt{\Delta^2 + g^2}\right)^2}} \begin{pmatrix} \Delta \pm \sqrt{\Delta^2 + g^2} \\ g \end{pmatrix}. \quad (3)$$

The new wavefunction for hybrid mode is a linear combination of $|o\rangle$ and $|a\rangle$ with the complex amplitude $A_o$ and $A_a$, respectively. The eigenvectors in Eq. (3) are presented on the representation $\{|o\rangle, |a\rangle\}$. Conventionally, we rewrite the wavefunction as:

$$\emptyset = \begin{pmatrix} e^{i\delta} A_1 \\ A_2 \end{pmatrix} = [A_o \begin{pmatrix} -1 \\ 1 \end{pmatrix} + A_a \begin{pmatrix} 1 \\ 1 \end{pmatrix}]. \quad (4)$$

It links the obscure wavefunction hybridization to the two observed parameters, including the amplitudes $A_i$ (i=1,2) and phase $\delta$.

### III. Simulation results and discussion

To confirm the validity of Eqs. (2) and (4) on the magnon–magnon hybrid modes, we performed micromagnetic simulations on a SAF with in-plane uniaxial magnetic anisotropy (UMA). In our previous work, we observed an anisotropic magnon–magnon coupling in this particular type of SAF [15]. The maximum coupling strength was observed when the magnetic field was applied at an angle of 45° relative to the UMA direction [25]. In this study, we focus on investigating the magnon hybrid nature in this typical scenario, with a specific emphasis on wavefunction hybridization. The parameters of the SAF and further details regarding the micromagnetic simulation can be found in the Supplementary Material. As discussed earlier, the hybridization of magnon wavefunctions leads to intricate dynamics, which is observed in the obtained dynamical signals. Fig. 1(c) shows a typical pair of dynamical traces presented by their polar-components m$_{1z}$ and m$_{2z}$ under the magnetic field $H$=300 Oe. Since the broadband excitation, both modes are trigged. The traces were fitted by two damped sine waves $m_{iz}(t) = A_{i-}e^{-t/\tau_{i-}} sin(2\pi f_- t + \psi_{i-}) + A_{i+}e^{-t/\tau_{i+}} sin(2\pi f_+ t + \psi_{i+})$ to extract dynamical parameters of m$_{iz}$ (i=1,2 for top and bottom layers). $A_{i\pm}$, $\tau_{i\pm}$, and $\psi_{i\pm}$ correspond to the characteristic amplitudes, decay times, and phases, respectively; $f$ presents the resonance frequencies; and + and – present the upper and lower branches, respectively. Then, the oscillation behavior of m$_{1z}$ and m$_{2z}$ decomposed into two oscillations with different resonance frequencies. Simultaneously, the linewidth was obtained as $\Delta f_\pm = 1/2\pi\tau_\pm$ .

The resonance frequencies and linewidths of the two branches are presented in Figs. 2(a) and 2(b), respectively. An evident anticrossing of the dispersion curves confirms the presence of magnon–magnon coupling. The coupling strength, denoted as g, is determined as 0.43 GHz. Moreover, an intercrossing of the linewidths has been observed, which aligns with the predictions of Eq. (2) when $f_a = f_o$. This finding implies that even when categorizing the modes as AM-like and OM-like, the linewidths still coincide. This phenomenon indicates that the hybridization enhances the dissipation rate in one mode while suppressing it in the other. To quantify Eq. (2), we consider the resonance frequency and linewidth of the SAF without K$_u$ (uniaxial anisotropy) as the bare values, denoted as $f$ and $\Delta f$ , respectively (for a comprehensive explanation, please refer to the Supplementary Material). These bare values are plotted and marked as the calculation without g in Fig. 2. By including the coupling strength g=0.43 GHz, we calculate the resonance frequencies and linewidths, which are presented in Fig. 2. The calculated resonance frequencies closely match the real part of the eigenfrequencies (Eq. (2)), except for field strengths below 120 Oe. This discrepancy arises from neglecting the uniaxial anisotropy field when considering the bare $f$. The trend of the linewidths is also well reproduced by the imaginary part of the eigenfrequencies. The slight differences (as indicated by the scaling factor in Fig. 2(b)) can be attributed to the neglect of the uniaxial anisotropy field, which slightly modifies the effective field and consequently affects the dissipation rate (Eqs. S3–S4). Nonetheless, the validity of Eq. (2) is confirmed, thereby establishing the two-coupled band model for characterizing and quantifying the magnon-magnon hybrid mode.

The eigenvectors were calculated using Eq. (3), and then the amplitudes $A_i$ and phase $\delta$ were determined using Eq. (4) for comparison with the simulation results. These parameters are in good agreement with the micromagnetic simulation results. The amplitude ratios $A_{1+}/A_{2+}$ for $f_+$ branch and $A_{2-}/A_{1-}$ for $f_-$ branch are plotted in Fig. 3(a), respectively. As the magnetic field increases, the ratios initially decrease and then increase. The minimum ratio approaches zero. A ratio value less than one indicates that $m_1$ dominates $f_-$ branch, while $m_2$ dominates $f_+$ branch. Fig. 3(b) illustrates the phase shift $\delta = \psi_2 - \psi_1$ for the two branches as a function of the magnetic field. With increasing field strength, $\delta$ changes from 0° to 180° for $f_-$ branch, while it changes from 180° to 0° for $f_+$ branch. This evolution of $\delta$ indicates that one branch transitions from being AM-like to OM-like, while the other branch transitions from being OM-like to AM-like. According to Eqs. (3) and (4), the presence of $\Delta f_a \neq \Delta f_o$ results in a complex value for $e^{i\delta}$, indicating that $\delta$ is neither 0° nor 180°. Therefore, the phase shift $\delta$ is associated with the non-Hermitian characteristics of wavefunction hybridization [25]. Overall, the hybridization of magnon wavefunctions leads to asymmetric precessional amplitudes and a phase shift that deviates from 0° and 180°.

In the context of a two-band model, the eigenvector $(A_o, A_a)^T$ possesses a two-fold symmetry and can be mapped to a pseudospin vector on a Bloch sphere [30–32]. We can express $(A_o, A_a)^T$ as $(cos\frac{\theta}{2}, e^{i\varphi}sin\frac{\theta}{2})^T$ with $\theta = 2arctan(abs(\frac{A_o}{A_a}))$ and $\varphi = \arg(\Delta \pm \sqrt{\Delta^2 + g^2})$, while $\varphi$ represents the phase shift between |o> and |a> in a hybrid mode. The value of $\varphi$ is depicted in Fig. S2 (please refer to the Supplementary Material). For the $f_+$ branch, φ takes a small but non-zero value, whereas for the $f_-$ branch, it is close to but not exactly −180°. Similar to the phase $\delta$, the complex value of $e^{i\varphi}$ reveals that phase $\varphi$ also possesses a non-Hermitian characteristic. We designate the state |o> and |a> as the north and south poles of the Bloch sphere, respectively. The two eigenvectors correspond to a pair of pseudospin vectors on the magnonic Bloch sphere. This pseudospin vector ($\theta, \varphi$) characterizes the wavefunction |h> as a superposition state of |o> and |a>: |h>=$cos\frac{\theta}{2}$|o> +$e^{i\varphi}sin\frac{\theta}{2}$|a>. Fig. 4 displays the evolution of the calculated pseudospin vector on the Bloch sphere as a function of the

external field. With increasing field strength, the pseudospin vector of the $f_+$ branch gradually transforms from the north pole to the south pole, whereas the opposite occurs for the $f_-$ branch. Therefore, the magnetic field serves as an effective parameter for tuning the hybridization.

The aforementioned discussions successfully quantify the modulation of resonance frequency, linewidth, precession amplitude, and phase, thus confirming the validity of the two-band model and wavefunction hybridization in describing the magnon hybrid state. Moreover, the hybridization presents a distinct departure from the explanation of mutual spin pumping for linewidth suppression or enhancement [21,22]. Further investigations are required to distinguish between these two different explanations. The asymmetry of precession amplitudes is proposed as an alternative picture of symmetry-breaking of the wavefunction, which is the key issue to understanding and introducing magnon–magnon coupling in SAF. The amplitude asymmetry can be experimentally characterized using techniques, such as electrically-detected ferromagnetic resonance (FMR) [33]. Additionally, the relative phase between two magnetic layers can be observed clearly through time-resolved X-ray magnetic circular dichroism, with well-designed samples [34]. The pseudospin vector on the magnonic Bloch sphere not only aids in characterizing the hybrid mode but also offers a means to tune the magnon state for quantum technology applications.

## IV. Conclusions

In summary, we propose a non-Hermitian two coupled band model to describe the wavefunction hybridization in magnon–magnon hybrid modes. We select in-phase and out-of-phase excitations as the basis wavefunctions, and the magnon–magnon coupling couples them to form a hybrid mode. Comparing our model with micromagnetic simulation results on a symmetrical SAF with UMA, we successfully replicate the resonance frequencies, linewidths, and reconstruct the wavefunction, including the phases and amplitudes for the two layers. The observed intercrossing of linewidths between the two magnon bands is attributed to the wavefunction hybridization. It results in a mixing of the wavefunctions of the AM and OM, leading to the mixing of the linewidths. Furthermore, we analyze the magnon hybridization in a magnonic Bloch sphere as a function of the external field. Characterizing hybrid modes is a key aspect in understanding and developing hybrid quantum systems. Our work provides a unique perspective for understanding and characterizing magnon hybridization in magnon-based quantum hybrid systems.

## Acknowledgements

This work is supported by the National Natural Sciences Foundation of China (Grant Nos. 12174427, 51871235).

Fig.1

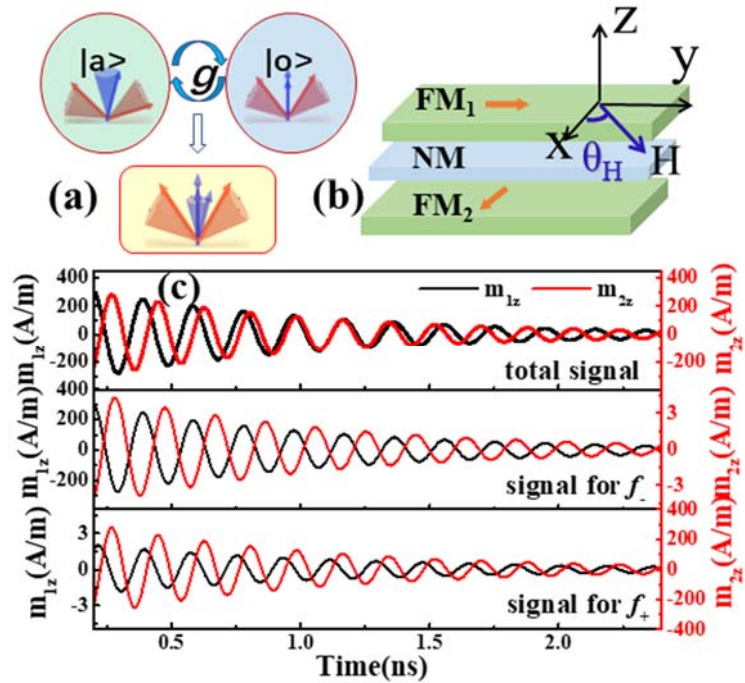

Fig.1 (a) (color online) The diagram of two basis magnon wavefunction |a>, |o>, and the wavefunctions hybridization mediated by magnon-magnon coupling to form a magnon-magnon hybrid mode, respectively. (b) The diagram of magnetization configuration in a synthetic antiferromagnet. The easy axis is along x-axis. The magnetic field is applied in xy plane with $\theta_H$ azimuth angle along the x-axis direction. (c) The magnetization oscillations of $m_{1z}$ (black) and $m_{2z}$ (red) as the total signal, the signal for lower and upper frequencies, respectively.

Fig.2

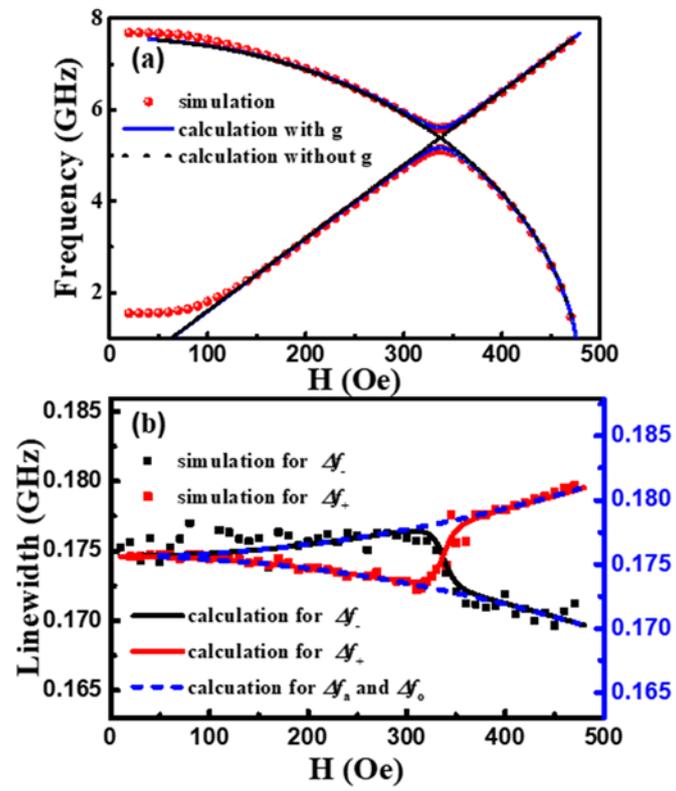

Fig.2 (a) The resonance frequencies and (b) frequency-linewidths for the two magnon branches at the function of external field. The simulation values are obtained from the micromagnetic simulation and the calculation values are obtained from Eq. (2), respectively.

FIg.3

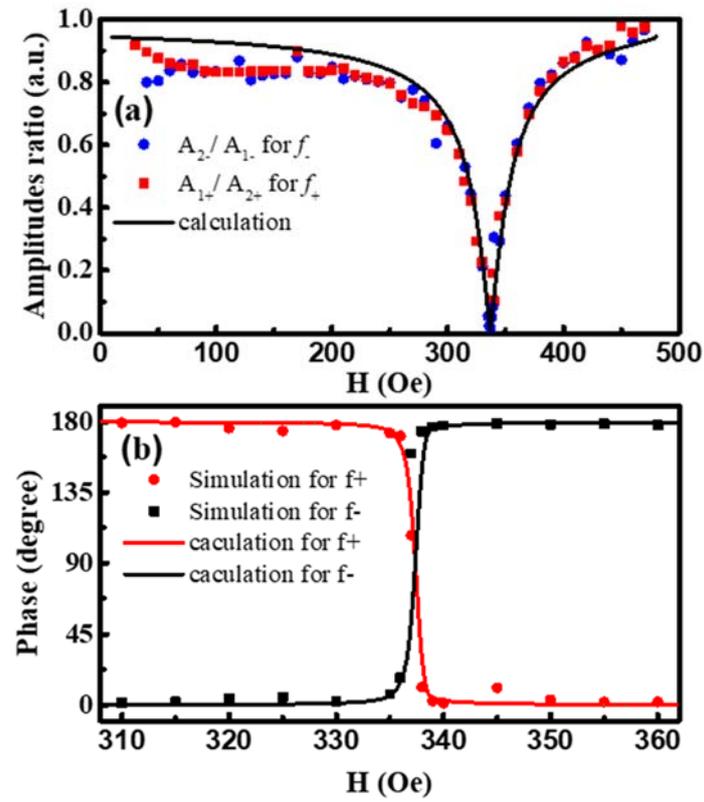

Fig.3 (a)The amplitude ratio and (b) the phase shift δ between two layer of the hybrid magnon mode at the function of external field. The simulation values are obtained from the micromagnetic simulation and the calculation values are obtained from Eq. (4), respectively.

Fig. 4

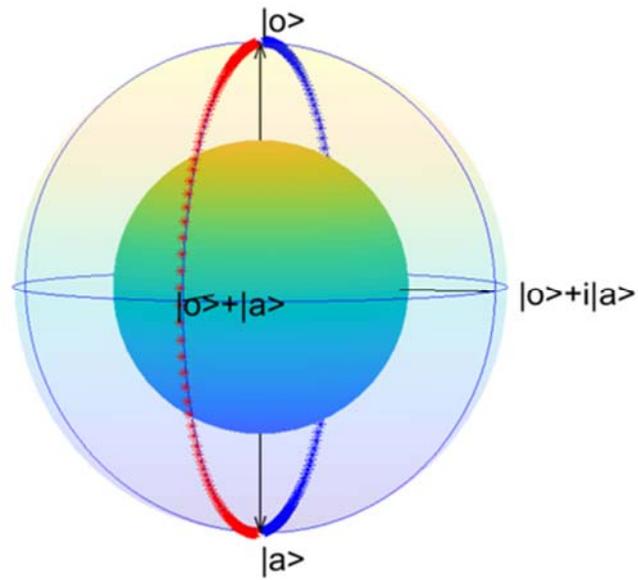

Fig. 4 (color online) Hybridization pseudo-spin state $|h\rangle = cos\frac{\theta}{2}|0\rangle + e^{i\varphi} sin\frac{\theta}{2}|1\rangle$ of the magnonic Bloch sphere. The red dots correspond to $f_+$ branch and the blue dots correspond to $f_-$ branch, respectively.